\documentclass[showpacs,10pt,twocolumn,prl]{revtex4-1}

\usepackage{amsmath}
\usepackage{amssymb}
\usepackage{graphicx}
\usepackage{amssymb}
\usepackage{graphics}
\usepackage{epsfig}
\usepackage{CJK}
\usepackage{color}

\begin{document}

\begin{CJK*}{GBK}{Song}
\title{Three-dimensional Ising Ferrimagnetism of Cr-Fe-Cr trimers in FeCr$_2$Te$_4$}
\author{Yu Liu,$^{1}$ R. J. Koch,$^{1}$  Zhixiang Hu,$^{1,2}$ Niraj Aryal,$^{1}$ Eli Stavitski,$^{3}$ Xiao Tong,$^{4}$ Klaus Attenkofer,$^{3}$ E. S. Bozin,$^{1}$ Weiguo Yin,$^{1}$ and C. Petrovic$^{1,2}$}
\affiliation{$^{1}$Condensed Matter Physics and Materials Science Department, Brookhaven National Laboratory, Upton, New York 11973, USA\\
$^{2}$Department of Physics and Astronomy, Stony Brook University, Stony Brook, New York 11790, USA\\
$^{3}$National Synchrotron Light Source II, Brookhaven National Laboratory, Upton, New York 11973, USA\\
$^{4}$Center for Functional Nanomaterials, Brookhaven National Laboratory, Upton, New York 11973, USA}
\date{\today}

\begin{abstract}
We carried out a comprehensive study of magnetic critical behavior in single crystals of ternary chalcogenide FeCr$_2$Te$_4$ that undergoes a ferrimagnetic transition below $T_c$ $\sim$ 123 K. Detailed critical behavior analysis and scaled magnetic entropy change indicate a second-order ferrimagentic transition. Critical exponents $\beta = 0.30(1)$ with $T_c = 122.4(5)$ K, $\gamma = 1.22(1)$ with $T_c = 122.8(1)$ K, and $\delta = 4.24(2)$ at $T_c$ $\sim$ 123 K suggest that the spins approach three-dimensional Ising ($\beta$ = 0.325, $\gamma$ = 1.24, and $\delta$ = 4.82) model coupled with the attractive long-range interactions between spins that decay as $J(r)\approx r^{-4.88}$. Our results suggest that the ferrimagnetism in FeCr$_2$Te$_4$ is due to itinerant ferromagnetism among the antiferromagnetically coupled Cr-Fe-Cr trimers.
\end{abstract}
\maketitle
\end{CJK*}

\section{INTRODUCTION}

Ternary ACr$_2$X$_4$ (A = transition metal, X = S, Se, and Te) exhibit a variety of magnetic and electronic properties. The family includes metallic CuCr$_2$X$_4$ and semiconducting Hg(Cd)Cr$_2$Se$_4$ ferromagnets \cite{Takeshi,HM,Men}, semiconducting Fe(Mn)Cr$_2$S$_4$ ferrimagnets \cite{Yang,Tsu,OHG}, and insulating ZnCr$_2$S$_4$ antiferromagnet \cite{Hem}. The FeCr$_2$X$_4$ compounds show competing spin-orbit and exchange interactions \cite{Be}. FeCr$_2$S$_4$ is ferrimagnetic (FIM) insulator below $T_c$ = 165 K, shows a crossover transition from insulator to metal near $T_c$ and colossal magnetoresistance behavior \cite{Tsurkan,Tsurkan1,Ramirez}. FeCr$_2$Se$_4$ is an insulating antiferromagnet (AFM) with $T_N$ = 218 K and ferrimagnetic with a small magnetic moment of 0.007 $\mu_B$ below 75 K \cite{Min,Snyder,OK}. It should be noted that FeCr$_2$Se$_4$ crystallizes in the Cr$_3$S$_4$-type monoclinic structure described within $C2/m$ space group, in contrast to the cubic spinel-type of FeCr$_2$S$_4$. However, FeCr$_2$S$_4$ and FeCr$_2$Se$_4$ have similar electronic structure with nearly trivalent Cr$^{3+}$ and divalent Fe$^{2+}$ states \cite{Kang}. The magnetic moments of Cr ions are antiparallel to those of Fe ions in FeCr$_2$X$_4$, and there is strong hybridization between Fe $3d$-states and X p-states \cite{Kang}.

FeCr$_2$Te$_4$ has not been studied much presumably due to the difficulty in sample preparation \cite{Andre,Valiev,Yadav}. Demeaux et al. first grew the single crystals of FeCr$_2$Te$_4$ \cite{Andre}. The crystal structure was reported as a defective NiAs-type within $P6_3/mmc$ space group, where Fe and Cr occupy the same site with alloying ratio of 1 : 2 and a net occupancy of 0.75 \cite{Andre}. In contrast, Valiev et al. reported that FeCr$_2$Te$_4$ crystalizes in a CoMo$_2$S$_4$-type structure within $I2/m$ space group, in which Fe and Cr are octahedrally coordinated by six Te \cite{Valiev}. Recently, a series of polycrystals FeCr$_2$Se$_{4-x}$Te$_x$ were synthesized \cite{Yadav}. Substitution with Te gradually suppresses the AFM order of FeCr$_2$Se$_4$, and leads to a short range ferromagnetic (FM) cluster metallic state in polycrystal FeCr$_2$Te$_4$ \cite{Yadav}. In order to study the intrinsic physical property, high quality single crystal is required.

In this work, we successfully fabricated single crystals of FeCr$_2$Te$_4$ and performed a comprehensive study of the structural and magnetic properties. Our analysis of criticality around $T_c$ indicates that FeCr$_2$Te$_4$ displays the 3D-Ising behavior, with the magnetic exchange distance decaying as $J(r)\approx r^{-4.88}$. Our first principles calculations suggest that the ferrimagnetism in FeCr$_2$Te$_4$ stems from the itinerant ferromagnetism among the antiferromagnetically coupled Cr-Fe-Cr trimers. Since transition-metal chalcogenides represent model systems for exploring local structure-related relationship between the broken symmetry and d-orbital magnetism \cite{Bozin}, detailed local stucture investigation of this system would be highly desirable and would bring important new insights.

\section{EXPERIMENTAL DETAILS}

Single crystals of FeCr$_2$Te$_4$ were fabricated by melting stoichiometric mixture of Fe (99.99\%, Alfa Aesar) powder, Cr (99.95\%, Alfa Aesar) powder, and Te (99.9999\%, Alfa Aesar) pieces. The starting materials were vacuum-sealed in a quartz tube, heated to 1200 $^\circ$C over 12 h, slowly cooled to 900 $^\circ$C at a slow rate of 1 $^\circ$C/h, and then quenched into iced water. The single crystal x-ray diffraction (XRD) data were taken with Cu $K_{\alpha}$ ($\lambda=0.15418$ nm) radiation of a Rigaku Miniflex powder diffractometer. In order to obtain more comprehensive crystallographic information, the powder XRD measurements were performed at the PDF beamline (28-ID-1) at National Synchrotron Light Source II (NSLS II) at Brookhaven National Laboratory (BNL) using Perkin Elmer image plate detector. The setup utilized x-ray beam with a wavelength of 0.1666 {\AA}, and the sample to detector distance of 1.25 m, as calibrated using a Ni standard. Sample was cooled with an Oxford Cryosystems 700 cryostream using liquid nitrogen. Raw data were integrated and converted to intensity vs. scattering angle using the software pyFAI \cite{Kieffer}. The average structure was assessed from raw powder diffraction data using the General Structure Analysis System II (GSAS-II) software package \cite{Toby}. The elemental analysis was performed using energy-dispersive x-ray spectroscopy (EDS) in a JEOL LSM-6500 scanning electron microscope (SEM). The x-ray absorption spectroscopy (XAS) measurements were performed at 8-ID beamline of the NSLS II (BNL) in a fluorescence mode. The x-ray absorption near edge structure (XANES) and extended x-ray absorption fine structure (EXAFS) spectra were processed using the Athena software package. The AUTOBK code was used to normalize the absorption coefficient, and separate the EXAFS signal, $\chi(k)$, from the atom-absorption background. The extracted EXAFS signal, $\chi(k)$, was weighed by $k^2$ to emphasize the high-energy oscillation and then Fourier-transformed in $k$ range from 2 to 10 {\AA}$^{-1}$ to analyze the data in $R$ space. The x-ray photoelectron spectroscopy (XPS) experiment was carried out in an ultrahigh vacuum system with base pressures $< 5\times10^{-9}$ Torr, equipped with a hemispherical electron energy analyzer (SPECS, PHOIBOS 100) and a twin anode x-ray source (SPECS, XR50). Al $K_{\alpha}$ (1486 eV) radiation was used at 13 kV and 30 mA. The angle between the analyzer and the x-ray source was 45$^\circ$ and photoelectrons were collected along the sample surface normal. The XPS spectra were analyzed and deconvoluted using the Casa XPS software. The dc/ac magnetic susceptibility were measured in Quantum Design MPMS-XL5 system. The applied field ($H_a$) was corrected as $H = H_a-NM$, where $M$ is the measured magnetization and $N$ is the demagnetization factor. The corrected $H$ was used for the analysis of magnetic entropy change and critical behavior.

\section{RESULTS AND DISCUSSIONS}

\begin{figure}
\centerline{\includegraphics[scale=0.7]{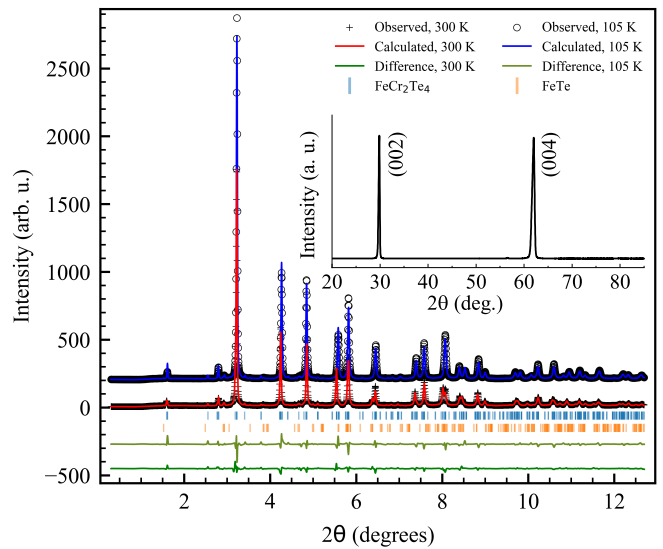}}
\caption{(Color online) Synchrotron powder x-ray diffraction (XRD) data and structural model refinements. The data are shown by (+) and ($\circ$), structural model fit at 300 K and 105 K is shown by red and blue solid line, respectively. The difference curves are given by green solid lines, offset for clarity. The vertical tick marks represent Bragg reflections of the $I2/m$ space group and up to 4\% of FeTe impurity. Inset shows single crystal XRD pattern of FeCr$_2$Te$_4$ at room temperature.}
\label{XRD}
\end{figure}

\begin{figure}
\centerline{\includegraphics[scale=0.35]{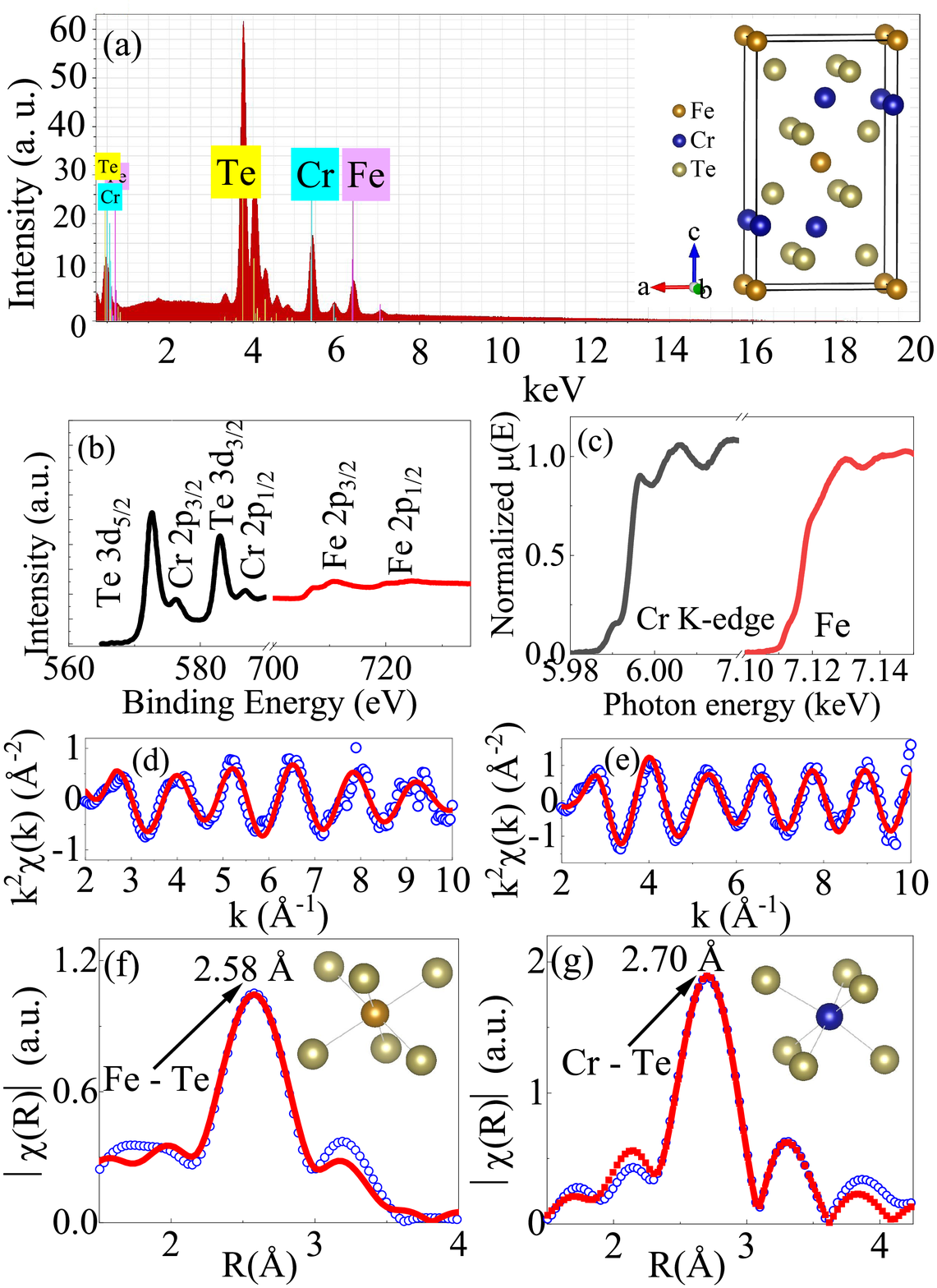}}
\caption{(Color online) (a) Crystal structure and results of EDS analysis on FeCr$_2$Te$_4$. Room-temperature Fe-2p, Cr-2p, and Te-3d core-level XPS (b) and normalized Cr and Fe K-edge XANES spectra (c). Fe and Cr K-edge EXAFS oscillations (d,e) and Fourier transform magnitudes (f,g) of EXAFS data measured at room temperature. The experimental data are shown as blue symbols alongside the model fit plotted as red line. Corresponding first coordination shell of Fe and Cr are shown in the insets.}
\label{XRD}
\end{figure}

\begin{table}
\caption{\label{tab}Average and local structural parameters extracted from the powder XRD and the EXAFS spectra of FeCr$_2$Te$_4$. CN is coordination number based on crystallographic value, R is interatomic distance, and $\sigma^2$ is Debye Waller factor.}
\begin{ruledtabular}
\begin{tabular}{lllll}
  \multicolumn{1}{c}{} &\multicolumn{2}{c}{300 K} &\multicolumn{2}{c}{105 K}\\
  \hline
  \multicolumn{1}{c}{$a$ ({\AA})} &\multicolumn{2}{c}{6.822(2)} &\multicolumn{2}{c}{6.825(1)}\\
  \multicolumn{1}{c}{$b$ ({\AA})} &\multicolumn{2}{c}{3.938(1)} &\multicolumn{2}{c}{3.943(1)}\\
  \multicolumn{1}{c}{$c$ ({\AA})} &\multicolumn{2}{c}{11.983(5)} &\multicolumn{2}{c}{11.867(5)}\\
  \multicolumn{1}{c}{$\beta$ ($^\circ$)} &\multicolumn{2}{c}{90.00(5)} &\multicolumn{2}{c}{90.01(15)}\\
  \hline
  atom & site & $x$ & $y$ & $z$  \\
  \hline
  Fe & $2a$ & 0 & 0 & 0  \\
  Cr & $4i$ & 0.001(3) & 0 & 0.2541(8)  \\
  Te1 & $4i$ & 0.328(4) & 0 & 0.3726(4)  \\
  Te2 & $4i$ & 0.339(4) & 0 & 0.8785(4)  \\
  \hline
  bond & CN & R ({\AA}) & $\Delta$R ({\AA}) & $\sigma^2$ ({\AA}$^2$)\\
  \hline
  Cr-Te1 & 2 & 2.58 & 0.01 & 0.001 \\
  Cr-Te2 & 1 & 2.67 & 0.01 & 0.001 \\
  Cr-Te3 & 1 & 2.80 & 0.11 & 0.003 \\
  Cr-Te4 & 2 & 2.86 & 0.11 & 0.003 \\
  Fe-Te1 & 4 & 2.69 & 0.36 & 0.02 \\
  Fe-Te2 & 2 & 2.75 & 0.24 & 0.02
\end{tabular}
\end{ruledtabular}
\end{table}

In the single-crystal XRD pattern (inset to Fig. 1), only $(00l)$ peaks were observed. Synchrotron powder XRD pattern of pulverized crystal of FeCr$_2$Te$_4$ can be well fitted by using a monoclinic structure with the $I2/m$ space group (Fig. 1), confirming main phase of FeCr$_2$Te$_4$ with less than 4\% FeTe impurity. The determined lattice parameters at 300 K are $a = 6.822(2)$ {\AA}, $b = 3.938(1)$ {\AA}, $c = 11.983(5)$ {\AA}, and $\beta = 90.00(5)^\circ$, close to the reported values \cite{Valiev,Yadav}. No structural transition was observed on cooling, showing a compression along $c$ axis with $c$ = 11.867(5) {\AA} and a slight expansion in the $ab$ plane with $a = 6.825(1)$ {\AA} and $b = 3.943(1)$ {\AA} down to 105 K (Table I).

The ratio of elements in the single crystal as determined by EDS is Fe : Cr : Te = 0.99(2) : 1.90(2) : 4.0(1) [Fig. 2(a)], and it is referred to as FeCr$_2$Te$_4$ throughout this paper. The information on the valence states of Fe, Cr and Te atoms can be obtained from the element core-level XPS spectra [Fig. 2(b)]: Fe$^{2+}$ ($2p_{3/2}$ $\sim$ 710 eV, $2p_{1/2}$ $\sim$ 723 eV), Cr$^{3+}$ ($2p_{3/2}$ $\sim$ 575 eV, $2p_{1/2}$ $\sim$ 586 eV), Te$^{2-}$ ($3d_{5/2}$ $\sim$ 572 eV, $3d_{3/2}$ $\sim$ 583 eV)]. Figure 2(c) shows the normalized Cr and Fe K-edge XANES spectra, in which a similar prepeak feature is observed, indicating similar local atomic environment for Fe and Cr atoms. The prepeak feature for Fe K-edge is somewhat weaker than that of Cr K-edge, suggesting a weaker lattice distortion in FeTe$_6$ when compared with CrTe$_6$. The edge features are close to the standard compounds with Cr$^{3+}$ and Fe$^{2+}$ oxidation states \cite{Ig,Ofuchi}, in line with the XPS result.

The local environment of Fe and Cr atoms is revealed in the EXAFS spectra of FeCr$_2$Te$_4$ measured at room temperature [Figs. 2(d) and 2(e)]. In a single-scattering approximation, the EXAFS can be described by \cite{Prins}:
\begin{align*}
\chi(k) = \sum_i\frac{N_iS_0^2}{kR_i^2}f_i(k,R_i)e^{-\frac{2R_i}{\lambda}}e^{-2k^2\sigma_i^2}sin[2kR_i+\delta_i(k)],
\end{align*}
where $N_i$ is the number of neighbouring atoms at a distance $R_i$ from the photoabsorbing atom. $S_0^2$ is the passive electrons reduction factor, $f_i(k, R_i)$ is the backscattering amplitude, $\lambda$ is the photoelectron mean free path, $\delta_i$ is the phase shift, and $\sigma_i^2$ is the correlated Debye-Waller factor measuring the mean square relative displacement of the photoabsorber-backscatter pairs. The corrected main peak in the Fourier transform magnitudes of Fe K-edge EXAFS around $R \sim 2.58$ {\AA} is clearly smaller than that of Cr K-edge EXAFS at $R \sim 2.70$ {\AA} [Figs. 2(f) and 2(g)]. The different local Fe-Te and Cr-Te bond lengths suggest that the Fe and Cr atoms might occupy different crystallographic sites, ruling out the possibility of NiAs-type structure with the same Fe/Cr sites. Then we focus on the first nearest neighbors of Fe and Cr atoms ranging from 1.5 to 3.5 {\AA}. The main peak corresponds to two Fe-Te bond lengths of 2.69 {\AA}  and 2.75 {\AA}, and four different Cr-Te bond lengths with 2.58 {\AA}, 2.67 {\AA}, and 2.80 {\AA}, 2.86 {\AA}, respectively, extracted from the model fits with fixed coordination number $CN$. The peaks above 3.25 {\AA} are due to longer Fe-Cr, Fe-Te, and Cr-Te bond distances, and the multiple scattering involving different near neighbours of the Fe/Cr atoms.

Figure 3(a) shows the temperature dependence of magnetization measured in out-of-plane field $\mu_0H$ = 0.1 T, in which $\chi$ increases with decreasing temperature and increases abruptly near $T_c$ due to the paramagnetic (PM)-FIM transition. The in-plane $\chi(T)$ [inset in Fig. 3(a)] is much smaller than that in out-of-plane field, indicating the presence of large magnetic anisotropy with easy $c$ axis. The average susceptibility $\chi_{ave} = (2/3)\chi_{ab}+(1/3)\chi_c$ from 150 to 300 K can be fitted by the Curie-Weiss law $\chi_{ave} = \chi_0 + C/(T-\theta)$ [Fig. 3(c)], which yields $\chi_0$ = 0.87(1) emu/mol-Oe, $C$ = 7.91(6) emu-K/mol-Oe [$\mu_{eff} = 7.95(9) \mu_B$ per formula unit], and $\theta$ = 126.6(2) K. The positive value of $\theta$ indicates dominance of ferromagnetic or ferrimagnetic exchange interactions in FeCr$_2$Te$_4$. For monoclinic FeCr$_2$Se$_4$ with a similar layered structure, the individual spins of Fe and Cr ions have AFM coupling along the $c$ axis with the distance of 2.956 {\AA}, while FM coupling along the $b$ axis with the distance of 3.617 {\AA} \cite{Adachi}. At low temperature, the FM interaction dominating over the AFM interaction results in a FIM ground state. For FeCr$_2$Te$_4$, the enhanced hybridization between $d$-orbital of Fe and Cr with $p$-orbital of Te plays an important role in magnetic coupling. Based on our first-principle calculation (see below), the FIM structure where Fe and Cr atoms have opposite spin orientations are the most stable state in FeCr$_2$Te$_4$, when compared with the FM and AFM structures, which needs further verification by neutron scattering experiment. The bifurcation between ZFC and FC curves [Fig. 3(a)] might be due to strong magnetic anisotropy and/or multidomain structure, which has also been observed in other long-range FM single crystals, such as U$_2$RhSi$_3$ \cite{Szlawska}. The magnetization loops of FeCr$_2$Te$_4$ for both field directions at $T$ = 2 K confirms a large magnetic anisotropy and easy $c$ axis [Fig. 3(c)]. The sudden jumps around $\mu_0H$ $\approx$ $\pm$ 1 T along the $c$ axis can be ascribed to the magnetic domain creeping behavior, i.e., the magnetic domain walls jump from one pinning site to another. Then we estimated the Rhodes-Wohlfarth ratio (RWR) for FeCr$_2$Te$_4$, which is defined as $P_c/P_s$ with $P_c$ obtained from the effective moment $P_c(P_c+2) = P_{eff}^2$ and $P_s$ is the saturation moment obtained in the ordered state \cite{Wohlfarth,Moriya,Takahashi}. RWR is 1 for a localized system and is larger in an itinerant system. Here we obtain RWR $\approx$ 1.69 for FeCr$_2$Te$_4$, indicating a weak itinerant character. To obtain the accurate Curie temperature $T_c$, out-of-plane ac susceptibility was measured at oscillating ac field of 3.8 Oe and frequency of 499 Hz. The single sharp peak in the real part $m^\prime(T)$ [Fig. 3(d)] gives the $T_c$ = 124 K.

\begin{figure}
\centerline{\includegraphics[scale=1]{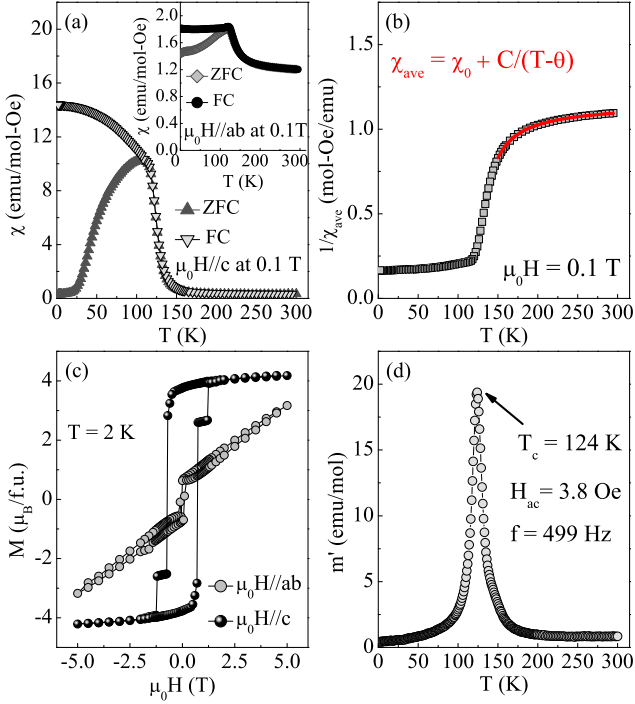}}
\caption{(Color online) Temperature-dependent dc magnetic susceptibility $\chi(T)$ in zero-field cooling (ZFC) and field cooling (FC) modes taken at $\mu_0H$ = 0.1 T for $\mathbf{\mu_0H\parallel c}$ (a) and $\mathbf{\mu_0H\parallel ab}$ (inset), respectively. (b) $1/\chi(T)$ taken at $\mu_0H$ = 0.1 T along with Curie-Weiss fit from 150 to 300 K. (c) Field dependence of magnetization for FeCr$_2$Te$_4$ measured at $T$ = 2 K. (d) Ac susceptibility real part $m'(T)$ measured with oscillating ac field of 3.8 Oe and frequency of 499 Hz. The chosen experimental parameters (field and frequency) allow for well defined moment and adequate frequency resolution.}
\label{MTH}
\end{figure}

\begin{figure}
\centerline{\includegraphics[scale=0.4]{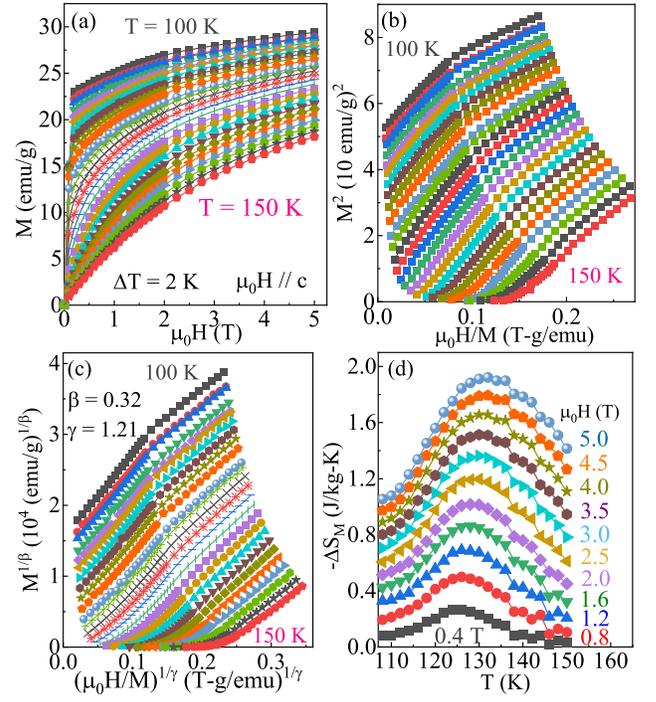}}
\caption{(Color online) (a) Typical initial isothermal magnetization curves from 100 K to 150 K with a temperature step of 2 K for FeCr$_2$Te$_4$ single crystal. (b) Arrott plot ($\beta = 0.5$, $\gamma = 1$) and (c) the modified Arrott plot with the optimum exponents $\beta = 0.32$ and $\gamma = 1.21$. (d) Temperature-dependent magnetic entropy change $-\Delta S_M(T)$ at various fields change.}
\label{KF}
\end{figure}

In the following we discuss the nature of the PM-FIM transition for FeCr$_2$Te$_4$. The magnetization isotherms along easy $c$ axis were measured at various temperatures in the vicinity of $T_c$ [Fig. 4(a)]. We first considered the well-known Arrott plot \cite{Arrott1}. From the Landau theory, the Arrott plot of $M^2$ vs $H/M$ should appear as parallel straight lines above and below $T_c$, and the line passes through the origin at $T_c$. It is clear that the mean field critical exponent does not work for FeCr$_2$Te$_4$, as illustrated by the set of curved lines shown in Fig. 4(b).

For a second-order phase transition, the spontaneous magnetization $M_s$ below $T_c$, the inverse initial susceptibility $\chi_0^{-1}$ above $T_c$, and the field-dependent magnetization $M(H)$ at $T_c$ are \cite{Stanley,Fisher,Lin}:
\begin{equation}
M_s (T) = M_0(-\varepsilon)^\beta, \varepsilon < 0, T < T_c,
\end{equation}
\begin{equation}
\chi_0^{-1} (T) = (h_0/m_0)\varepsilon^\gamma, \varepsilon > 0, T > T_c,
\end{equation}
\begin{equation}
M = DH^{1/\delta}, T = T_c,
\end{equation}
where $\varepsilon = (T-T_c)/T_c$ is the reduced temperature, and $M_0$, $h_0/m_0$ and $D$ are the critical amplitudes. In a more general case, the modified Arrott plot $(H/M)^{1/\gamma} = a\varepsilon + bM^{1/\beta}$ with self-consistent method was considered \cite{Kellner,Pramanik}. Figure 4(c) presents the final modified Arrott plot of $M^{1/\beta}$ vs $(H/M)^{1/\gamma}$ with $\beta = 0.32$ and $\gamma = 1.21$, showing a set of quasi-parallel lines at high field region. Then we extracted $\chi_0^{-1}(T)$ and $M_s(T)$ as the intercepts on the $H/M$ axis and the positive $M^2$ axis, respectively. The magnetic entropy change can be estimated using the Maxwell's relation \cite{Amaral}:
\begin{equation}
\Delta S_M(T,H) = \int_0^H \left[\frac{\partial M(T,H)}{\partial T}\right]_HdH.
\end{equation}
Figure 4(d) presents the calculated $-\Delta S_M$ as a function of temperature. The $-\Delta S_M$ shows a broad peak centered near $T_c$ and the peak value monotonically increases with increasing field. The maximum value of $-\Delta S_M$ reaches 1.92 J kg$^{-1}$ K$^{-1}$ with a field change of 5 T. There is a slight shift of $-\Delta S_M$ peak towards higher temperature with increasing field, which also excludes the mean field model \cite{Francoo}.

\begin{figure}
\centerline{\includegraphics[scale=1]{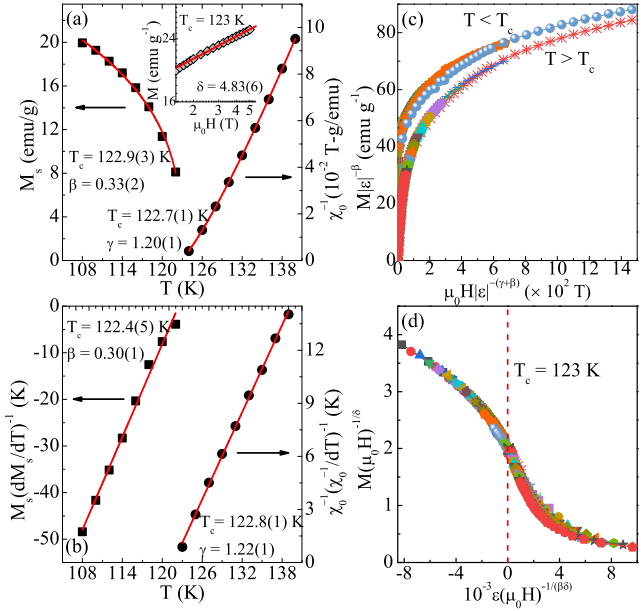}}
\caption{(Color online) (a) Temperature dependence of the spontaneous magnetization $M_s$ (left) and the inverse initial susceptibility $\chi_0^{-1}$ (right) with solid fitting curves. Inset shows log$M$ vs log$(\mu_0H)$ collected at $T_c$ = 123 K with linear fitting curve. (b) Kouvel-Fisher plots of $M_s(dM_s/dT)^{-1}$ (left axis) and $\chi_0^{-1}(d\chi_0^{-1}/dT)^{-1}$ (right axis) with solid fitting curves. (c) Scaled magnetization $m$ vs scaled field $h$ below and above $T_c$ for FeCr$_2$Te$_4$. (d) The rescaling of the $M(\mu_0H)$ curves by $M(\mu_0H)^{-1/\delta}$ vs $\varepsilon (\mu_0H)^{-1/(\beta\delta)}$.}
\label{renomalized}
\end{figure}

\begin{table*}
\caption{\label{tab1}Comparison of critical exponents of FeCr$_2$Te$_4$ with different theoretical models.}
\begin{ruledtabular}
\begin{tabular}{llllllll}
   & Reference & Technique & $T_{c-}$ & $T_{c+}$ & $\beta$ & $\gamma$ & $\delta$ \\
  \hline
  FeCr$_2$Te$_4$ & This work & Modified Arrott plot & 122.9(3) & 122.7(1) & 0.33(2) & 1.20(1) & 4.6(2)\\
  & This work & Kouvel-Fisher plot & 122.4(5) & 122.8(1) & 0.30(1) & 1.22(1) & 5.1(1) \\
  & This work &Critical isotherm  &   & &   &   & 4.83(6) \\
  3D Heisenberg & 28 & Theory & & & 0.365 & 1.386 & 4.8 \\
  3D XY & 28 & Theory & & & 0.345 & 1.316 & 4.81 \\
  3D Ising & 28 & Theory & & & 0.325 & 1.24 & 4.82 \\
  Tricritical mean field & 36 & Theory & & & 0.25 & 1.0 & 5.0
\end{tabular}
\end{ruledtabular}
\end{table*}

Figure 5(a) presents the extracted $M_s(T)$ and $\chi_0^{-1}(T)$ as a function of temperature. According to Eqs. (1) and (2), the critical exponents $\beta = 0.33(2)$ with $T_c = 122.9(3)$ K, and $\gamma = 1.20(1)$ with $T_c = 122.7(1)$ K, are obtained. The exponent $\beta$ describes the rapid increase of the order parameter below $T_c$. The exponent $\gamma$ describes how magnetic susceptibility diverges at $T_c$. Here the obtained result describes critical behavior of the net spontaneous magnetization that arises in ferrimagnet. In the Kouvel-Fisher (KF) relation \cite{Kouvel}:
\begin{equation}
M_s(T)[dM_s(T)/dT]^{-1} = (T-T_c)/\beta,
\end{equation}
\begin{equation}
\chi_0^{-1}(T)[d\chi_0^{-1}(T)/dT]^{-1} = (T-T_c)/\gamma.
\end{equation}
Linear fittings to the plots of $M_s(T)[dM_s(T)/dT]^{-1}$ and $\chi_0^{-1}(T)[d\chi_0^{-1}(T)/dT]^{-1}$ in Fig. 5(b) yield $\beta = 0.30(1)$ with $T_c = 122.4(5)$ K, and $\gamma = 1.22(1)$ with $T_c = 122.8(1)$ K. The third exponent $\delta$ can be calculated from the Widom scaling relation $\delta = 1+\gamma/\beta$ \cite{Widom}. From $\beta$ and $\gamma$ obtained with the modified Arrott plot and the Kouvel-Fisher plot, $\delta$ = 4.6(2) and 5.1(1) are obtained, respectively, which are close to the direct fit of $\delta$ = 4.83(6) taking into account that $M = DH^{1/\delta}$ at $T_c$ = 123 K [inset in Fig. 5(a)]. The obtained critical exponents of FeCr$_2$Te$_4$ are very close to the theoretically predicted values of 3D Ising model ($\beta$ = 0.325, $\gamma$ = 1.24, and $\delta$ = 4.82) (Table II).

Scaling analysis can be used to estimate the reliability of the obtained critical exponents and $T_c$. The magnetic equation of state in the critical region is expressed as
\begin{equation}
M(H,\varepsilon) = \varepsilon^\beta f_\pm(H/\varepsilon^{\beta+\gamma}),
\end{equation}
where $f_+$ for $T>T_c$ and $f_-$ for $T<T_c$, respectively, are the regular functions. Eq. (7) can be further written in terms of scaled magnetization $m\equiv\varepsilon^{-\beta}M(H,\varepsilon)$ and scaled field $h\equiv\varepsilon^{-(\beta+\gamma)}H$ as $m = f_\pm(h)$. This suggests that for true scaling relations and the right choice of $\beta$, $\gamma$, and $\delta$, scaled $m$ and $h$ will fall on universal curves above $T_c$ and below $T_c$, respectively. As shown in Fig. 5(c), all the data collapse on two separate branches below and above $T_c$, respectively. The scaling equation of state takes another form,
\begin{equation}
\frac{H}{M^\delta} = k\left(\frac{\varepsilon}{H^{1/\beta}}\right),
\end{equation}
where $k(x)$ is the scaling function. From the above equation, all the data should also fall into a single curve. This is indeed seen [Fig. 5(d)]; the $M(\mu_0H)^{-1/\delta}$ vs $\varepsilon (\mu_0H)^{-1/(\beta\delta)}$ experimental data collapse into a single curve and the $T_c$ locates at the zero point of the horizontal axis. The well-rescaled curves confirm the reliability of the obtained critical exponents and $T_c$.

Furthermore, it is important to discuss the nature as well as the range of magnetic interaction in FeCr$_2$Te$_4$. In a homogeneous magnet the universality class of the magnetic phase transition depends on the exchange distance $J(r)$. In renormalization group theory analysis the interaction decays with distance $r$ as $J(r) \approx r^{-(3+\sigma)}$, where $\sigma$ is a positive constant \cite{Fisher1972}. The susceptibility exponent $\gamma$ is:
\begin{multline}
\gamma = 1+\frac{4}{d}\left(\frac{n+2}{n+8}\right)\Delta\sigma+\frac{8(n+2)(n-4)}{d^2(n+8)^2}\\\times\left[1+\frac{2G(\frac{d}{2})(7n+20)}{(n-4)(n+8)}\right]\Delta\sigma^2,
\end{multline}
where $\Delta\sigma = (\sigma-\frac{d}{2})$ and $G(\frac{d}{2})=3-\frac{1}{4}(\frac{d}{2})^2$, $n$ is the spin dimensionality \cite{Fischer}. When $\sigma > 2$, the Heisenberg model is valid for the 3D isotropic magnet, where $J(r)$ decreases faster than $r^{-5}$. When $\sigma \leq 3/2$, the mean-field model is satisfied, expecting that $J(r)$ decreases slower than $r^{-4.5}$. For the 3D-Ising model with $d$ = 3 and $n$ = 1, $\sigma$ = 1.88 is obtained, leading to spin interactions $J(r)$ decaying as $J(r)\approx r^{-4.88}$. This calculation suggests that the spin interaction in FeCr$_2$Te$_4$ is close to the 3D Ising localized-type coupled with a long-range ($\sigma = 1.88$) interaction, in line with its weak itinerant character. Meanwhile, the correlation length ($\xi$) correlates with the critical exponent $\nu$ ($\nu = \gamma/\sigma$), where $\xi = \xi_0[(T-T_c)/T_c]^{-\nu}$. It gives that $\nu$ = 0.64(1) and $\alpha$ = 0.08 ($\alpha= 2 - \nu d$).

\begin{figure}
\centerline{\includegraphics[scale=1]{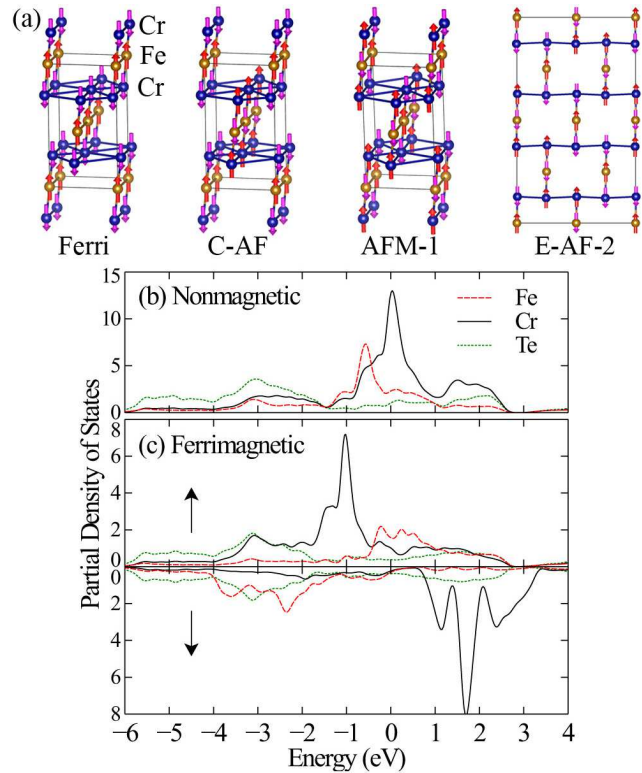}}
\caption{(Color online) (a) The antiferromagnetic structures
used in the first-principles total energy calculations. Atom-resolved density of states (b) in the nonmagnetic state and (c) in the ferrimagnetic state where the Cr and Fe atoms have opposite spin orientations, which are shown in the ratio of two Cr ions to one Fe ion.}
\label{Fig:DFT}
\end{figure}

\begin{table}
\caption{The first-principles total energy (in meV) per formula unit of different magnetic patterns as shown in Fig.~\ref{Fig:DFT}(a) using the 300 K and 105 K experimental structures. In the E-AF-2 pattern, a Cr atom is antiferromagnetically and ferromagnetically aligned with the closest and farthermost of its six nearest Cr neighbors, respectively; the opposite holds in the E-AF-1 pattern. FM means the ferromagnetic configuration.}
\begin{tabular}{c|ccccc}
\hline\hline
Structure & Ferri & FM & C-AF & E-AF-1 & E-AF-2 \\ \hline
300 K   &  8    & 90 & 197 & 31 & 82 \\
105 K   & $0$   & 56 & 171 & 72 & 32 \\
\hline\hline
\end{tabular}
\label{magnetic}
\end{table}

To get further insight into the magnetism, we performed first-principles calculations using density function theory. We applied the WIEN2K implementation \cite{wien2k} of the full potential linearized augmented plane-wave method in generalized-gradient approximation using the PBEsol functional \cite{PBEsol}. The basis size was determined by $R_\mathrm{mt}K_\mathrm{max} = 7$ and the Brillouin zone was sampled with
115 irreducible $k$ points to achieve energy convergence of 1 meV. As shown in Table~\ref{magnetic}, we found that for the experimental structure refined at 105~K, the ferrimagnetic state where the Cr and Fe atoms have opposite spin orientations is 56 meV per formula unit lower in total energy than the ferromagnetic phase and 32 meV lowered than the most stable antiferromagnetic structure (i.e., E-AF-2) as observed in FeCr$_2$Se$_4$ \cite{Hong}. A similar trend of the results holds for the the experimental structure refined at 300~K (Table~\ref{magnetic}). A weak easy $c$-axis anisotropy was obtained by inclusion of spin-orbit coupling in the calculations, namely the total energy per formula unit is lower by less than 1 meV for the magnetization along the $c$ axis than along the $a$ or $b$ axis. Thus, other sources of magnetic anisotropy such as dipole-dipole interaction are important. The calculated atom-resolved density of states (DOS) is shown in Figs.~\ref{Fig:DFT}(b) and \ref{Fig:DFT}(c). For the nonmagnetic case, the Cr-derived DOS has a sharp peak at the Fermi level [Figs.~\ref{Fig:DFT}(b)], suggesting a strong Stoner instability that yields an itinerant ferromagnetism in the Cr layers. This yields the splitting of about 2.8 eV between the spin-majority and spin-minority bands of Cr character with the dramatic reduction of the Cr-derived DOS at the Fermi level [Fig.~\ref{Fig:DFT}(c)], indicative of the localized spin picture for the Cr atoms. Whereas, the nonmagnetic Fe-derived DOS peaks at about $0.6$ eV below the Fermi level [Fig.~\ref{Fig:DFT}(b)] and experiences little reduction at the Fermi level upon entering the magnetic phase [Fig.~\ref{Fig:DFT}(c)]. We infer that the magnetism of the Fe ions is established via antiferromagnetic superexchange with the neighboring two Cr ions. The Cr magnetic moment within the atomic Muffin tins is about 2.88 $\mu_{B}$, which is close to the nominal Cr$^{3+}$ $S=3/2$ state. With the octahedral coordination, the splitting of the five $3d$ orbitals between the high-lying $e_g$ and low-lying $t_{2g}$ orbitals is substantial. The $S=3/2$ state of the Cr$^{3+}$ ion (i.e., $3d^3$ or $t_{2g}^3e_g^0$ electron configuration) means that the Cr $t_{2g}$ orbitals are half filled, rendering a vanishing orbital angular moment and a negligible spin-orbit coupling effect. The Fe magnetic moment is about 2.80 $\mu_{B}$, significantly deviated from the nominal high-spin Fe$^{2+}$ $S=2$ state, which indicates its dual characters with both localized spins and itinerant electrons. This reveals an interesting interplay of Cr and Fe electronic states, which allows the spin-majority bands of the system at the Fermi level to be of Fe character rather than Cr character [Fig.~\ref{Fig:DFT}(c)]. We thus picture the FIM in FeCr$_2$Te$_4$ as itinerant ferromagnetism among the antiferromagnetically coupled Cr-Fe-Cr trimers. The trimers centered at the Fe sites form a body-centered orthorhombic lattice of magnetic dipoles with effective moment of $2\mu_{B}S=4\mu_{B}$ [Fig.~\ref{MTH}(c)]. In addition to the effective Heisenberg exchange interaction, the $i$th trimer is coupled with its ten neighboring trimers [Fig.~\ref{Fig:DFT}(a)] via dipole-dipole interaction $\propto -(\mathbf{S}_i\cdot \mathbf{r}_{ij})(\mathbf{S}_j\cdot \mathbf{r}_{ij})/|\mathbf{r}_{ij}|^5$, which tends to align the magnetic moments along the bond direction $\mathbf{r}_{ij}=\mathbf{r}_{i}-\mathbf{r}_{j}$ where $j$ denotes one of the neighboring trimers and $\mathbf{r}_{i}$ is the spatial vector of the $i$th trimer. Since the body-centered orthorhombic structure of the trimers is substantially elongated along the $c$ axis, easy $c$-axis magnetic anisotropy  has the overall minimum deviation from the neighboring bond directions. We thus predict that the 3D Ising-like ferrimagnetism is sensitive to changes in the lattice structure, especially the tilting of the Cr-Fe-Cr trimers, which will be verified by future pressure experiments and computer simulations.

\section{CONCLUSIONS}

In summary, we systematically investigated structural and magnetic properties of stoichiometric FeCr$_2$Te$_4$ that crystallizes in the $I2/m$ space group. The second-order PM-FIM transition is observed at $T_c$ $\sim$ 123 K. The critical exponents $\beta$, $\gamma$, and $\delta$ estimated from various techniques match reasonably well and follow the scaling equation. The analysis of critical behavior suggests that FeCr$_{2}$Te$_{4}$ is a 3D-Ising system displaying a long-range exchange interaction with the exchange distance decaying as $J(r)\approx r^{-4.88}$. Combined experimental and theoretical analysis attributes the ferrimagnetism in FeCr$_2$Te$_4$ to itinerant ferromagnetism among the antiferromagnetically coupled Cr-Fe-Cr trimers. Follow-up studies of local atomic structure and magnetism using x-ray and neutron scattering as well as high-pressure methods will be of particular interest for more comprehensive understanding of this system.

\section*{Acknowledgements}
Work at BNL is supported by the Office of Basic Energy Sciences, Materials Sciences and Engineering Division, U.S. Department of Energy (DOE) under Contract No. DE-SC0012704. This research used the 28-ID-1 and 8-ID beamlines of the NSLS II, a U.S. DOE Office of Science User Facility operated for the DOE Office of Science by BNL under Contract No. DE-SC0012704. This research used resources of the Center for Functional Nanomaterials (CFN), which is a U.S. DOE Office of Science Facility, at BNL under Contract No. DE-SC0012704.

\end{document}